\documentclass[useAMS,usenatbib]{mn2e}

\usepackage{subfig}
\usepackage{graphicx}

\title[Cluster Galaxies Evolution in $\Lambda$CDM]{The Growth in Size and Mass of Cluster Galaxies since $z=2$}

\author[Laporte et al.]{
\parbox[t]{\textwidth}{
Chervin F. P. Laporte$^{1}$, Simon D. M. White$^{1}$, Thorsten Naab$^{1}$, Liang Gao$^{2,3}$
}
\\
$^{1}$ Max Planck Institute for Astrophysics, Karl-Schwarzschild-Strasse 
1, 85740 Garching, Germany\\
$^{2}$ Partner Group of the Max Planck Institute for Astrophysics, National Astronomical Observatories,\\ 
Chinese Academy of Sciences, Beijing, 100012, China\\
$^{3}$ 2Institute of Computational Cosmology, Department of Physics, University of Durham,Science Laboratories,\\
 South Road, Durham DH1 3LE\\
}
\begin{document}
\date{}
\pagerange{\pageref{firstpage}--\pageref{lastpage}} \pubyear{2011}
\maketitle
\label{firstpage}
\begin{abstract}
We study the formation and evolution of Brightest Cluster Galaxies starting from a $z=2$ population of quiescent ellipticals and following them to $z=0$. To this end, we use a suite of nine high-resolution dark matter-only simulations of galaxy clusters in a $\Lambda$CDM universe. We develop a scheme in which simulation particles are weighted to generate realistic and dynamically stable stellar density profiles at $z=2$. Our initial conditions assign a stellar mass to every identified dark halo as expected from abundance matching; assuming there exists a one-to-one relation between the visible properties of galaxies and their host haloes. We set the sizes of the luminous components according to the observed relations for $z\sim2$ massive quiescent galaxies. We study the evolution of the mass-size relation, the fate of satellite galaxies and the mass aggregation of the cluster central. From $z=2$, these galaxies grow on average in size by a factor 5 to 10 of and in mass by 2 to 3. The stellar mass growth rate of the simulated BCGs in our sample is of 1.9 in the range $0.3<z<1.0$ consistent with observations, and of 1.5 in the range $0.0<z<0.3$. Furthermore the satellite galaxies evolve to the present day mass-size relation by $z=0$. Assuming passively evolving stellar populations, we present surface brightness profiles for our cluster centrals which resemble those observed for the cDs in similar mass clusters both at $z=0$ and at $z=1$. This demonstrates that the $\Lambda$CDM cosmology does indeed predict minor and major mergers to occur in galaxy clusters with the frequency and mass ratio distribution required to explain the observed growth in size of passive galaxies since $z=2$. Our experiment shows that Brightest Cluster Galaxies can form through dissipationless mergers of quiescent massive $z=2$ galaxies, without substantial additional star formation.
\end{abstract}

\section{Introduction}
Brightest cluster galaxies (BCGs) form the massive end of the galaxy population. They are generally associated with old stellar populations, little star formation and large sizes \citep{vonderLinden2007,Bernardi2009}, except in strong cooling flows. In some clusters, BCGs are surrounded by a diffuse envelope of intracluster light. This additional light has been measured in a number of nearby clusters \citep{Gonzalez2005} as well as in stacks of BCGs from the Sloan Digital Sky Survey \citep{Zibetti2005}. There have been claims that the observed evolution of BCGs (e.g. \citealp{Collins2009,Stott2011}) disagrees with the predictions of semi-analytic models of galaxy formation \citep{Delucia2007}. These studies selected BCGs in high redshift clusters and compared them to the central galaxies of present-day clusters of the same X-ray luminosity, observing little change in the sizes and stellar masses. However, recent results from Lidman et al. (2012) seem to indicate less tension between the models and observations. It is important to note that the current semi-analytic models of galaxy formation do not predict surface brightness profiles in a self-consistent manner. Thus, any direct comparison of sizes in the models and in real galaxies should be made with caution. 

Observations of $z=2$ galaxies by \cite{Daddi2005} and \cite{Trujillo2007} revealed the presence of a population of massive quiescent galaxies with much smaller sizes than similar mass present-day ellipticals. This result has been confirmed by several other groups \citep{Dokkum2008,Newman2012}. Minor mergers \citep{Bezanson2009,Naab2009} have been proposed as the main mechanisms driving the recent size evolution of elliptical galaxies. Currently, it is unclear whether all such galaxies will grow into present-day ellipticals. \cite{Bernardi2009} proposed that some of these objects might be progenitors of today's BCGs.

\cite{White1976} and \cite{Ostriker1977} introduced galactic cannibalism as a possible mechanism to explain the formation of cD galaxies: a central galaxy gradually swallows its companions as dynamical friction brings them to the cluster centre. These deposit their stellar material primarily on the outskirts of the larger galaxy, helping it grow in size and mass. This phenomenon seems to fit well within the $\Lambda$CDM cosmogony where structure grows hierarchically. For example, semi-analytic models of galaxy formation find that BCGs form in a two-phase process: an initial collapse with rapid cooling and star formation at high redshift is followed by later growth through multiple dissipationless mergers of pre-existing progenitors \citep{Khochfar2006,Delucia2007}. A similar two-phase formation mechanism is also reported in hydrodynamical simulations of massive elliptical formation \citep{Naab2009,Oser2010,Feldmann2011}. Notably, \cite{Oser2012} show that compact massive ellipticals can grow onto the present-day mass-size relation through minor mergers. However, such simulations still produce galaxies that are too massive for the host haloes they inhabit so their quantitative applicability is in some doubt.

\cite{Dubinski1998} studied whether BCGs could form out of a population of spirals between redshift $z=3$ and $z=0$. His cosmologically motivated simulation showed that BCGs could potentially form through repeated dissipationless mergers of galaxies. This inspired later studies testing the collisionless merger hypothesis within cosmological simulations which provided new insights into the evolution of BCGs \citep{Ruszkowski2009} and into the origin of the intracluster light \citep{Rudick2006}. A persistent issue, however, is that the initial galaxies assumed appear imconsistent with recent high redshift observations. \cite{Laporte2012} showed that re-scaling the luminous components of the galaxies to bring them into better agreement with these observations changed the size growth of BCGs from $r\propto M$ to $r\propto M^{2}$ consistent with observational studies of massive quiescent galaxies at fixed number density \citep{Dokkum2010, Patel2012}

Here, we present a scheme which can weight particles in a high-resolution cosmological simulation of cluster formation to represent realistic stellar density profiles for the initial galaxies. Using this method we can assign to every halo of mass $M$ at a certain redshift $z_{i}$, a stellar mass $m_{*}$ according to an appropriate abundance matching relation, hereafter AMR \citep{Kravtsov2004, Vale2004, Moster2010,Guo2010, Behroozi2010, Behroozi2012, Moster2012}. This ensures a faithful representation of the high redshift luminosity function at $z=2$. We can then study the evolution of cluster galaxies from a population of galaxies consistent {\it both} with the luminosity function and with the mass-size relation observed at $z=2$. We aim to test the dissipationless merger hypothesis \citep{White1976, Ostriker1977, Dubinski1998}. The questions we want to address are the following: are BCGs special or simply the product of merging of normal cluster galaxies? Does star formation contribute substantially to the growth in mass of BCGs between $z=2$ and $z=0$? Do mergers explain why BCGs seem to lie off the present-day mass-size relation? Could BCGs have evolved from the observed high-redshift population of massive quiescent galaxies?

Section 2 presents the simulations we use and our method of generating stellar density profiles. In section 3, we study the properties of the BCGs and compare them with observations. In section 4 we study the evolution of BCGs and how it compares to the population of cluster ellipticals. We discuss the significance of our results in section 5 and conclude in section 6.

\section{Methods}
\subsection{Simulations}
We use a set of nine dark-matter-only zoom-in simulations of galaxy clusters from the {\it Phoenix} project \citep{Gao2012}. These are named Ph-A to Ph-I. The haloes were initially selected from the {\it Millennium Simulation} \citep{Springel2005b} and re-simulated with comoving softening length $\epsilon=0.3h^{-1}\, kpc$ and mass resolution $m_{p}\sim4-10\times10^{6} h^{-1} {\rm M_{\odot}}$. Details of the simulations are given in \cite{Gao2012}. The subhaloes were identified with the structure finder {\sc Subfind} \citep{Springel2001}. We also compute the potential of every particle in each subhalo at redshift $z=2$. For the rest of the discussion, our units are $\rm kpc$ and $\rm M_{\odot}$ for length and mass respectively.

\subsection{A Weighting scheme for cosmological dark matter simulations}
The weighting scheme presented here generalises that of \cite{Bullock2005} for direct use in cosmological dark matter simulations of structure formation. Cold dark matter (just like stars in a galaxy) is collisionless and its distribution function ({\sc df}) satistfies the collisionless Boltzmann equation (CBE). Provided that the {\sc df} of the dark matter in a halo is in a steady state, Jeans' theorem guarantees the existence of a distribution function of the form $f=f(I_{1}, I_{2}, I_{3})$, where $I_{1,2,3}$ are isolating integrals of the motion. In a triaxial potential a {\sc df} of the form $f=f(H)$, where $H=\frac{1}{2}v^{2}+\Phi$ is the Hamiltonian, can always be gerenated. Generally, dark haloes are not in a steady state; their lives are continuously shaped by accretion events such mergers and infall. However, in their study of the evolution of the {\sc df} of CDM haloes, \cite{Natarajan1997} showed that between merger events haloes are in phases of ``quasi-equilibrium'' within the virialised regions. Thus, provided we restrict ourselves to the virialised regions, Jeans' theorem can be invoked to generate a stellar {\sc df} of the form $f_{*}=f_{*}(E)$.

In order to generate a luminous stellar profile, we take each simulation particle of energy $E$ to simultaneously represent dark matter and stars in diferent amounts. through a weight function $\omega(E)=\frac{N_{*}(E)}{N(E)}=\frac{f_{*}(E)g(E)}{N(E)}$, where $N$ is the differential energy distribution, $g$ is the density of states and asterisks denote stellar quantities. %The differential energy distribution for the total profile of galaxies is naturally given by the simulation. 

We choose the stellar distribution to be represented by the \cite{Hernquist1990} profile:
\begin{equation}
\nu=\frac{a}{r\left(r+a\right)^3},
\end{equation}
where a is the scale radius which is related to the half-mass radius through $a=r_{e}/(\sqrt{2}+1)$. The half-mass radius in projection is related to the half mass radius through $R_{e}=1.33r_{e}$

We generate $f_{*}$, using Eddington's formula:
\begin{equation}
f_{*}(\mathcal{E})=\frac{1}{\sqrt{8}\pi^2} \int^{\mathcal{E}}_{0}\frac{d\Psi}{\sqrt{\mathcal{E}-\Psi}}\frac{d^2\nu(\Psi)}{d\Psi^2} + \frac{1}{\sqrt{\mathcal{E}}}\frac{d\nu}{d\Psi}\bigg|_{\Psi=0},
\end{equation}
 where  $\Psi=-\Phi+\Phi_{0}$ and $\mathcal{E}=-E+\Phi_{0}=\Psi-v^2/2$ are the relative potential and total energies respectively. The potential of an NFW profile is generally a reasonable description of the true potential and tends to zero in the limit $r$ goes to infinity thus we set $\Phi_{0}=0$.

The density of states is given by:
\begin{equation}
g(E)=(4\pi)^2\int^{r_{E}}_{0}r^{2}\sqrt{2(E-\Phi(r))}dr.\\
\end{equation}

Equations (2) and (3) are only valid for spherical systems, however provided an educated guess, they can also be applied to triaxial ones. Indeed, authors in the past have already used the same formalism to study the distribution function of dark haloes \citep{Natarajan1997}. To compute the $\frac{{\rm d}^{2}\nu}{{\rm d}\Phi^{2}}$ term in Eddington's formula, we have to relate $\Phi$ to $\nu$ monotonically as $\Phi(r)$ is multivalued in a triaxial potential. For this, we simply approximate $\Phi(r)= <\Phi(\mathbf{r})>$. However, in the definition of the energy, we retain the actual value of the potential associated with every particle. This ensures the generation of a profile with equidensity surfaces which follow the potential.

Figure 1 illustrates the kind of profile one can generate within the $z=2$ haloes of the Phoenix simulation. We have checked for the stability of our method by evolving a live dark halo for 150 Myr. We further looked for isolated galaxies in haloes which have not experienced any stellar mass growth between $z=2$ and $z=0$. We present such an example in Figure 2. This halo has grown in dark matter mass by a factor of two but its stellar mass profile has remained unchanged over 10 Gyr. However, the half-light radius (defined as the radius of a sphere containing half the total stellar mass) has increased by 30 percent. Although this is not a substantial problem for studying the evolution of BCGs, where the change in radius is much larger and primarily caused by material deposited on the outside of the galaxy \citep{Laporte2012} it warns us that our experiment is close to the resolution limit for studying the size growth of galaxies which do not accrete much stellar mass. Considering the large period of time over which this test has been carried out, we see that two-body relaxation processes are not the driving force behind the size growth we observe for the most massive galaxies.

\begin{figure}
\includegraphics[width=0.5\textwidth,trim=10mm 15mm 0mm 10mm,clip]{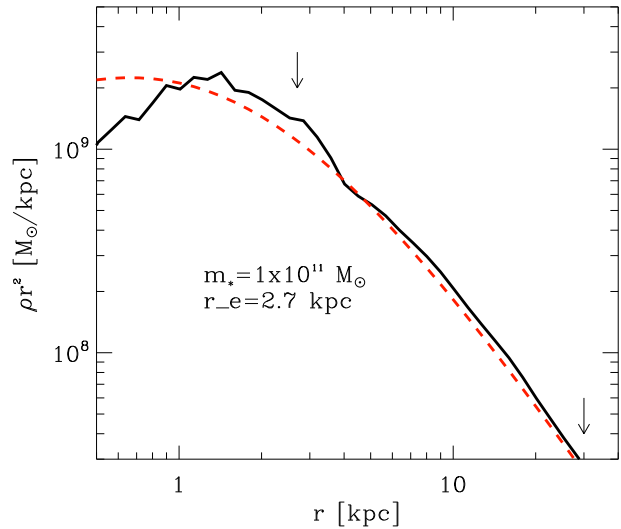}
\caption{Example $\rho r^{2}$ profile for a typical compact stellar profile generated using our weighting scheme. The dip is due to low particle numbers in the region below 1 kpc. Arrows mark the radii containing 50 and 95 percent of the stellar mass.}
\end{figure}

\begin{figure}
\includegraphics[width=0.5\textwidth,trim=0mm 0mm 0mm 0mm,clip]{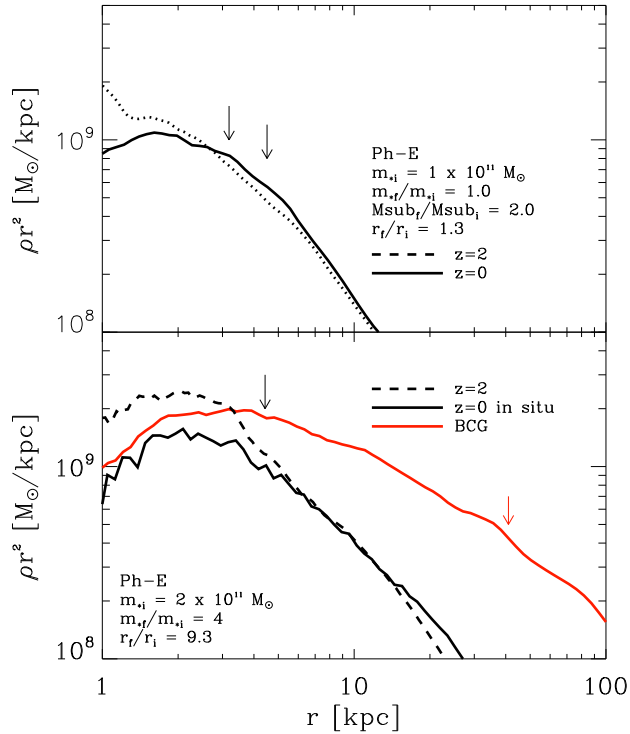}
\caption{{\it Top}:$\rho r^{2}$ profile for a galaxy which has evolved from $z=2$ (dashed line) to $z=0$ (solid line) without experiencing mergers or interactions with other haloes populated with stars. The host dark matter halo has grown in mass and half-mass radius by a factor of 2. This does not, however, translate in significant size growth of the stellar component. The arrows mark the radii enclosing 50 percent of the mass of the galaxies. {\it Bottom}: $\rho r^{2}$ profiles for a BCG most massive progenitor (dashed black line), and final $z=0$ BCG (red line). We also show the distribution of the stars in the BCG most massive progenitor at $z=0$ (solid black line). The BCG grew by a factor of four in stellar mass and nine in size. The black and red arrows show the radii enclosing 50 percent of the mass of the most massive BCG progenitor at $z=2$ and BCG at $z=0$ respectively. Clearly, the size growth observed for these galaxies is not due to two-body relaxation processes.}
\end{figure}

\subsection{Initial Conditions}

For our initial conditions, we choose to represent a population of galaxies consistent with the observed stellar mass function at $z=2$, which we relate to $\Lambda$CDM haloes in the simulations through recent AMR results \citep{Moster2010} which are consistent with results from \cite{Moster2012} even at $z=2$. We do not include the scatter in these relations, as the scatter for the mass-size relation is already large. We further require that the sizes of our model galaxies are consistent with those observed for massive quiescent galaxies $z=2$. For our purposes, we choose the mass-size relation as parametrised by \cite{Williams2010} and note that it is very similar to that of \cite{Newman2012} (see the 3rd panel of their Figure 8).

At $z=2$, a few haloes are undergoing mergers, but because these are a small fraction of the total we populate, we normally ignore them. However, if a particularly massive halo hosting a $\sim 10^{11} M_{\odot}$ galaxy is undergoing a merger we generate its stellar distribution function at an earlier snapshot. Depending on the state of the halo it is sometimes hard to generate stellar density profiles that perfectly match the original target, but as long as the galaxies are within the scatter of the observational relation we consider this good enough. For the rest of our discussion of the structural properties of ellipticals in clusters we will focus on objects which have masses above $7\times10^{10} \rm M_{\odot}$, which still leaves us with a sample of 156 galaxies at $z=2$. We follow, the subsequent evolution of galaxies by tracking the particle weights as the simulations evolve.

\section{Structural Properties of BCGs}

As a first test, it is interesting to check whether the total stellar mass of the final merger remnant agrees with expectations from the BCG luminosity function at $z=0$. Figure 3 shows the stellar-to-halo mass relation for the galaxies at $z=2$ and for cluster centrals and satellite galaxies at $z=0$.  As expected, many cluster galaxies have their dark matter haloes stripped, moving them horizontally in the $m_{*}-M_{halo}$ plane. Some of the cluster galaxies have a stellar mass deficit which can be explained by the lack of star formation in our experiment. These galaxies lie below the Moster et al. (2010) relation at $z=0$. Some cluster satellite galaxies however grow to stellar masses in agreement with the relation at $z=0$.  Moreover, the final BCGs occupy a region that is in good agreement with expectations of AMRs at $z=0$. This success depends on the assumption we made about the progenitor galaxies at redshift $z=2$ and on the hierarchical growth of the clusters according to the $\Lambda$CDM paradigm. At face value, this strongly supports the hypothesis that BCGs form from dissipationless mergers of galaxies that were already in place at $z=2$. This is in line with the galactic cannibalism picture originally formulated by \cite{White1976} and \cite{Ostriker1977}: BCGs {\it predominantly} grow through the later assembly of already existing galaxies. This idea agrees with the more recent studiy of \cite{Delucia2007} who find using their semi-analytic model that $80$ percent of the stars ending up in BCGs are already formed at $z=3$. With this in mind, we now ask whether any of the structural properties of our BCGs closely resemble those of known cluster central galaxies.

\begin{figure}
\includegraphics[width=0.5\textwidth,trim=0mm 0mm 0mm 0mm,clip]{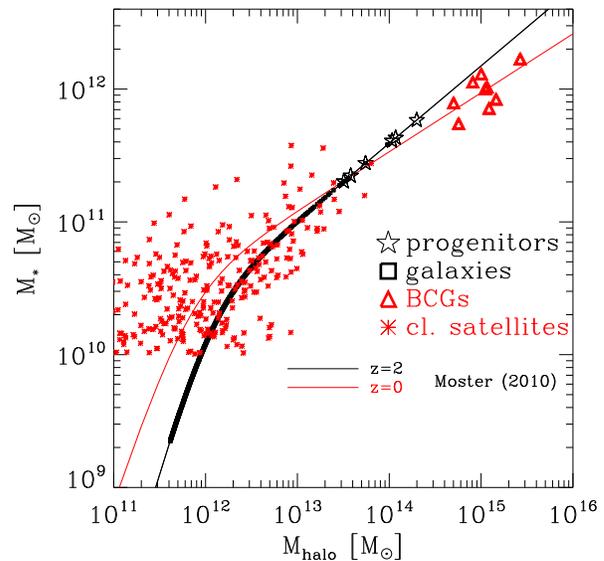}
\caption{Stellar-to-halo mass relations at z=2 and z=0 (black and red lines respectively). The black stars represent the most massive progenitors of BCGs at z=2 and the squares represent all the haloes populated with stars in the initial conditions. The red triangles are the BCGs at z=0 and the red crosses represent cluster satellite galaxies.}
\end{figure}
\subsection{Surface Brightness and Density Profiles}

\begin{figure}
\includegraphics[width=0.5\textwidth,trim=0mm 0mm 0mm 0mm,clip]{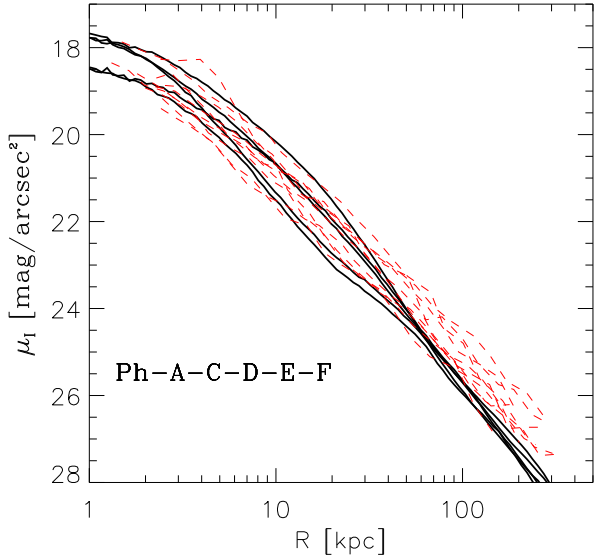}
\caption{Surface brightness profiles for BCGs at $z=0$ derived assuming passive evolution between $z=2$ and $z=0$. Overplotted in red are surface brightness profiles for nearby galaxy clusters from Gonzalez (2005). The simulated surface brightness profiles have a sligthly steeper fall-off at large radii compared to the observed ones.}
\end{figure}

Recent studies of massive quiescent galaxies at $z=2$, \citep{Williams2010, Dokkum2008} adopt a Kroupa IMF and solar metallicity when interpreting their photometric data and find a stellar population age of 1 Gyr. Our experiment assumes that the stellar populations evolve passively (the collisionless merger hypothesis), so we determine the mass-to-light ratios at later times using the values computed by \cite{Maraston2005}. We note that \cite{Newman2012} use Salpeter IMFs to derive their stellar masses, however due to the large scatter in the mass-size relation this is a minor issue. In fact, the two sets of data agree at $z=2$. 

In Figure 4, we present surface brightness profiles for BCGs at $z=0$. This is done by taking 50 random projections of the individual galaxies. Surface brightness profiles taken from \cite{Gonzalez2005} are overlaid in red. These authors measured the surface brightness profiles of 24 nearby BCGs in the I-band. The clusters they used were in the redshift range $0.03<z<0.1$. We find a reasonable match between the observed and simulated light distributions both in shape and in normalisations. However, we still note that at large radii our simulated galaxies have systematically slightly steeper fall-offs in their surface brightness profile.

Some authors have claimed that BCGs are already well in place at high redshift $z=1.0$ and that they evolve little thereafter \citep{Collins2009,Stott2011}. We compare the surface brightness profiles of some of our relaxed cluster BCGs to those of compiled by \cite{Stott2011} at $z=1.0$ in Figure 5. These authors deduced a stellar population formation age at $z=3$, similar to that is found for quiescent $z=2$ galaxies. The observations were performed in the {\it HST} F850LP band, but we compute our surface brightness profiles assuming a mass-to-light ratios in the SDSS z band which matches closely the F850LP band. We add the $(1+z)^4$ surface dimming correction for direct comparison with their data. Given the uncertainties in the IMF and the different band zero point, the agreement between the two is still reasonable and encouraging.

\begin{figure}
\includegraphics[width=0.5\textwidth,trim=0mm 0mm 0mm 0mm,clip]{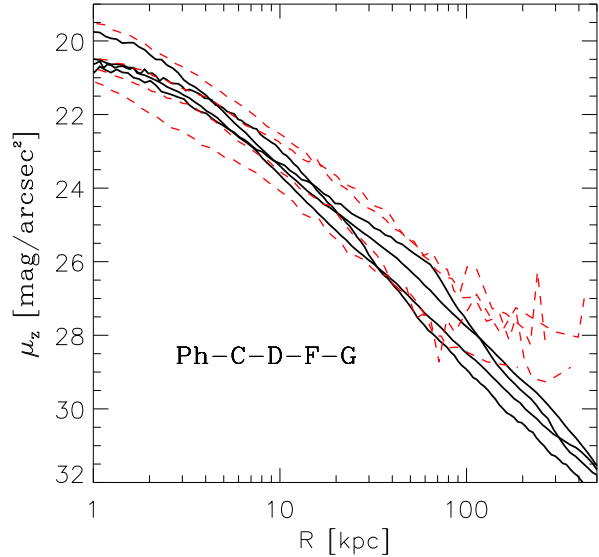}
\caption{Simulated surface brightness profiles in the SDSS z-band for BCGs at $z=1$ against projected radius. Overplotted in red are the observed surface brightness profiles in the F850LP HST band of for clusters of similar masses taken from the sample of Stott et al. (2011)}
\end{figure}

We also present the evolution in density profiles from $z=2$ to $z=0$ in Figure 6, separating the components into in-situ and accreted components. As previous studies have shown the size evolution is predominantly driven by adding stellar mass to the outskirts of the galaxies \citep{Naab2009, Oser2012, Laporte2012, Hilz2012a}.

\begin{figure*}
\includegraphics[width=0.9\textwidth,trim=0mm 0mm 0mm 0mm,clip]{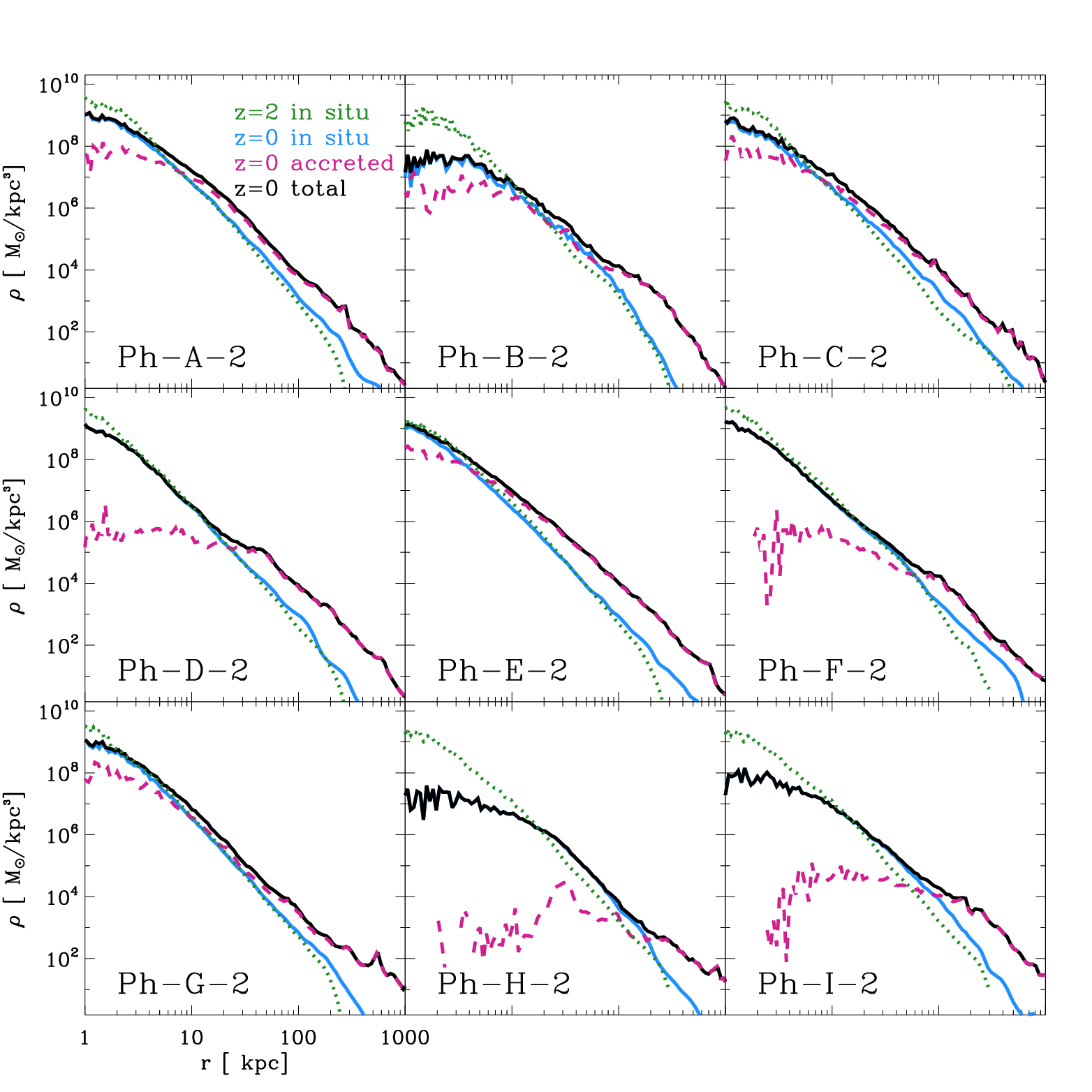}
\caption{Density profiles of BCGs for the nine cluster simulations. We separate in-situ components at $z=2$ (for the most massive progenitor) and $z=0$ (green and blue respectively) and accreted component at $z=0$ (magenta), the final BCG density profile is shown in black. Note that haloes B, H, I are unrelaxed at z=0 making their analysis cumbersome.}
\end{figure*}

\section{Evolution of BCGs and Ellipticals in Clusters}
%Our experiment has already demonstrated that the Brightest Cluster Galaxies form through collisionless mergers of already existing $z=2$ compact quiescent massive galaxies, reproducing their present-day properties: surface brightness, the extended envelope of stars at large radii, stellar-to-dark matter ratios in agreement with the latest AMRs at $z=0$.

Recently, \citeauthor{Lidman2012} (2012, hereafter L12), presented new results on the evolution of the stellar mass in BCGs using samples of high, intermediate and low redshift clusters. We compare their data with ours in Figure 7 looking at cluster mass versus BCG stellar mass at three different redshifts. Our BCGs lie within the scatter of their data, although generally towards the massive end in stellar mass. However, we show this is also partly due to the way L12 derive their stellar masses. For some of our BCGs, we used de Vaucouleurs profile fits within an aperture of 30 kpc to derive stellar masses. For high Sersic indices (e.g. $n=8$) up to half of the total stellar mass of the BCG can be missed. We indicate the shifts such systematics can cause for a few of our galaxies in Figure 7, showing that this might bring them in better agreement with the data. On the whole, L12 observe a stellar mass growth rate of 1.8 between $0.2<z<0.9$ for their sample. This is consistent with that found for our simulated galaxies: we observe a stellar mass-growth rate of $1.9$, $1.5$ and 2.6 between $0.3<z<1.0$, $0.0<z<0.3$, and $0<z<1$ respectively. The stellar mass growth since $z<1$ reported here is consistent with results from \cite{Delucia2007, Tonini2012}

\begin{figure}
\includegraphics[width=0.5\textwidth,trim=0mm 0mm 0mm 0mm,clip]{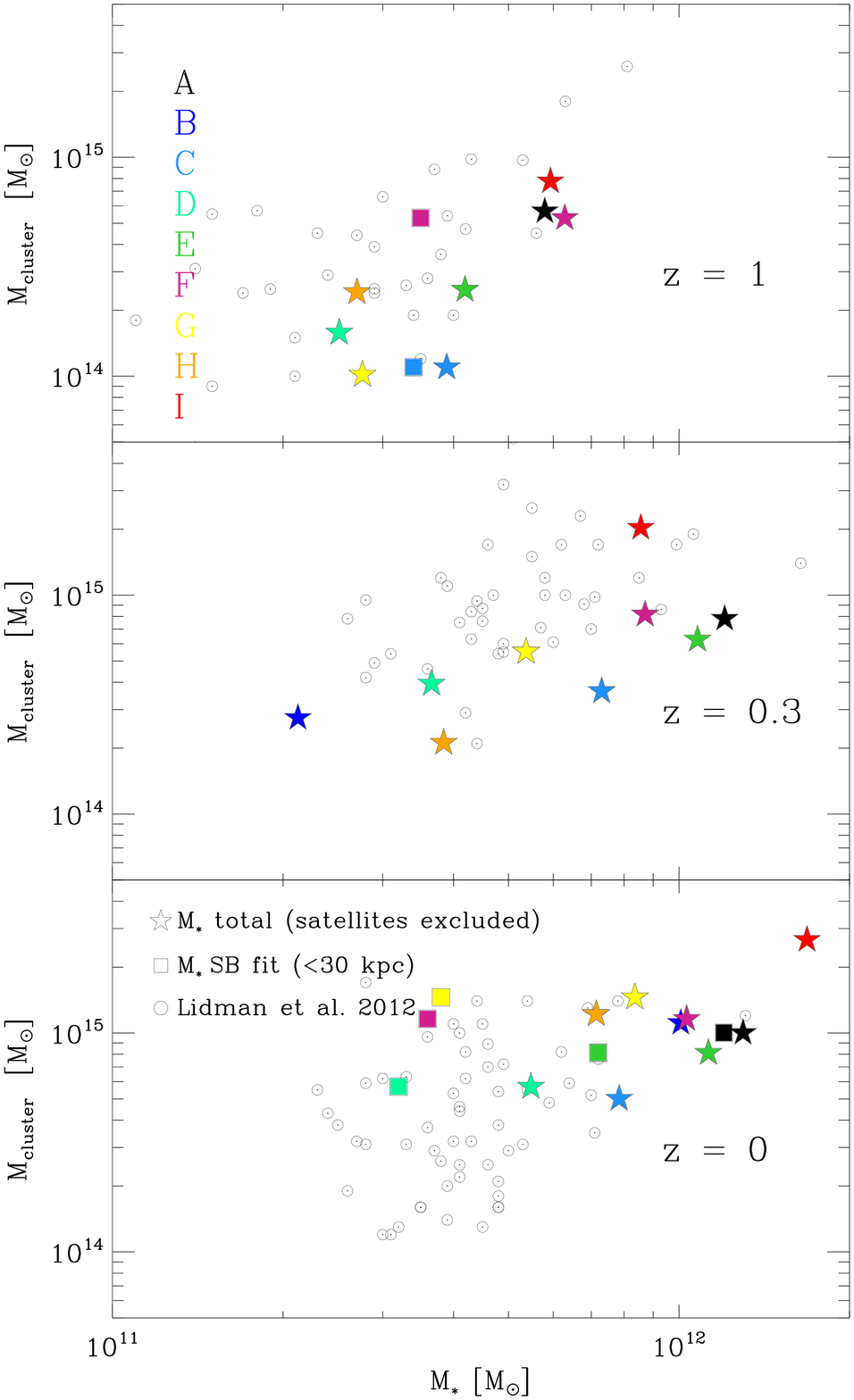} 
\caption{Cluster mass against BCG stellar mass at different redshifts. {\it Top:} High-$z$ BCG sample ($0.8<z<1.5$) and the Phoenix BCGs at $z=1.0$ (coloured stars). {\it Middle:} Intermediate-$z$ L12 BCGs ($0.2<z<0.5$) and Phoenix BCGs at $z=0.3$. {Bottom:} Low-$z$ ($z<0.2$) L12 BCGs and Phoenix BCGs. At $z=0$, we show for some BCGs the  stellar mass predicted by fitting a de Vaucouleurs profile within a 30 kpc aperture. For large indices (n=8 or n=10) this can lead to an underestimate of the true stellar mass of up to a factor of two. Models and observations are consistent at all $z$ given the large scatter in the observations.}
\end{figure}

We now turn to a comparison of the evolution in mass and size of the BCGs and that of the cluster population of ellipticals. We proceed by identifying all the $z=2$ haloes which end up in the final clusters at $z=0$ (defined as the major FOF group) and track them over redshift using the trees constructed from the {\sc subfind} data at $\sim 50$ intermediate outputs using individual particle IDs to match subhaloes in neighbouring outputs. As in \cite{Ruszkowski2009}, we compute three-dimensional spherical stellar mass profiles using logarithmic bins of widths $\Delta=0.1$ and measure their 3D half-mass radii. This is shown in Figure 8. We visually inspected the individual density profiles of cluster galaxies checking their Hernquist profile fits given the parameters determined in the previous step, namely $M$ and $a$. However, for BCGs, as their profile can sometimes be more extended (with Sersic profiles reaching $n=8-10$) than a simple de Vaucouleurs law, these fits  were often poor. Furthermore, we note that some clusters are out of equilibrium at various times which complicates the analysis of the central galaxy, this is the case for Ph-A-B-H-I at $z=1$ and for the Ph-B-H-I clusters at $z=0$.

Comparing panels, we see that BCGs evolve more rapidly than other galaxies and that different BCGs evolve at different rates. While some of the BCGs experience rapid growth in stellar mass by $z=1$ (e.g. Ph-A and Ph-H), others see most evolution between $z=0.3$ and $z=0$ (e.g. Ph-B, Ph-D Ph-G). There are also BCGs that show almost no significant evolution in stellar mass or size between $z=0.3$ and $z=0$ (e.g. Ph-A, Ph-E). Furthermore, some BCGs still suffer major mergers between $z=0.3$ and $z=0$, this is the case for Ph-I, which doubles in stellar mass. Such late phases of major merging are observed \citep{Liu2009,Brough2011}. We also see a growing mass gap between BCGs and other cluster ellipticals as we get closer to $z=0$. 

%This diversity in BCG evolution which already emerges out of nine cluster simulations should warn us of conclusions derived by studies with limited samples \cite{Nelson2002,Stott2010, Stott2011}. %{\bf need to add more} This also produces a substantial changes in sizes at relatively different times.

Our result confirms that of \cite{Ruszkowski2009}: BCGs tend to get off the mass-size relation of local ellipticals because of their much higher merger rate. Whether all BCGs should lie off the normal relation is harder to ascertain with a sample of only nine galaxy clusters. We find that at $z=0$, our cluster-ellipticals and some our BCGs lie on the mass-size relation derived by \cite{Hyde2009}. For comparison we also show in Figure 7 the \cite{Shen2003} relation which is shallower, a consequence of using Petrosian-based quantities which introduces a bias for objects of large Sersic index.

It is interesting to also look at the amount of mass deposited through mergers and disruption of satellites. We define a merger event with the BCG when half of the stellar mass of a progenitor galaxy gets incorporated within the first subhalo of the FOF group (defined as the halo inhabited by the BCG). Anything below this threshold we called ``diffuse'' stellar accretion from stripped satellites. We list in Table 1 the amount by which each type contributes to the mass aggregation for each BCGs. This table shows that for the various BCGs not all the mass accretion came from mergers but also some from stripped galaxies or galaxies in the process of merging. This amount of diffuse mass deposition, defined in Table 1 as the percentage of the final BCG mass diffusely accreted (i.e. $f_{diffuse}=(m_{acc}-m_{merger})/(m_{0}+m_{acc})$, where $m_{0}$ is the mass of the in-situ component of the most massive BCG progenitor),  can vary between five percent (Ph-C) and thirty percent (Ph-G and Ph-H) of the total accreted stellar mass. Furthermore, these values of accreted stellar mass are in agreement with the trends found in \cite{Moster2012}. We also present the number of mergers each BCGs experienced through their lifetime and list their ratios in Table 2. Although our definition of merger is somewhat arbitrary it clearly shows that BCGs go through a mixture and succession of major (1:1 - 1:2) and minor merging events (1:3 - 1:10) both of which are supported by observations of BCGs (e.g. \citealp{Liu2009,Edwards2012}). Furthermore, we see that the clusters for which we have identified fewer mergers have also a higher fraction of diffuse stellar accretion through stripped satellites: this is the case for clusters Ph-F, Ph-G and Ph-H. Although, there is a mixture of types of mergers, in the cases where minor merging has been predominantly acting (Ph-B, Ph-C, Ph-D, Ph-F, Ph-I), the numbers and ratios are in agreement with those found by the ad hoc simulations of \cite{Hilz2012b} to be needed to make the sizes of compact ellipticals grow to those of present day ones. 
 
%Minor mergers have been proposed as a way of puffing up $z=2$ compact ellipticals in order for them to grow to the present-day observed sizes. These claims have been supported by simple virial arguments \cite{Bezanson2009}, cosmological simulations \cite{Naab2009, Oser2011} as well as detailed collisionless merger studies of initially compact ellipticals \cite{Hilz2012}. 

\begin{figure*}
\includegraphics[width=0.9\textwidth,trim=0mm 0mm 0mm 0mm,clip]{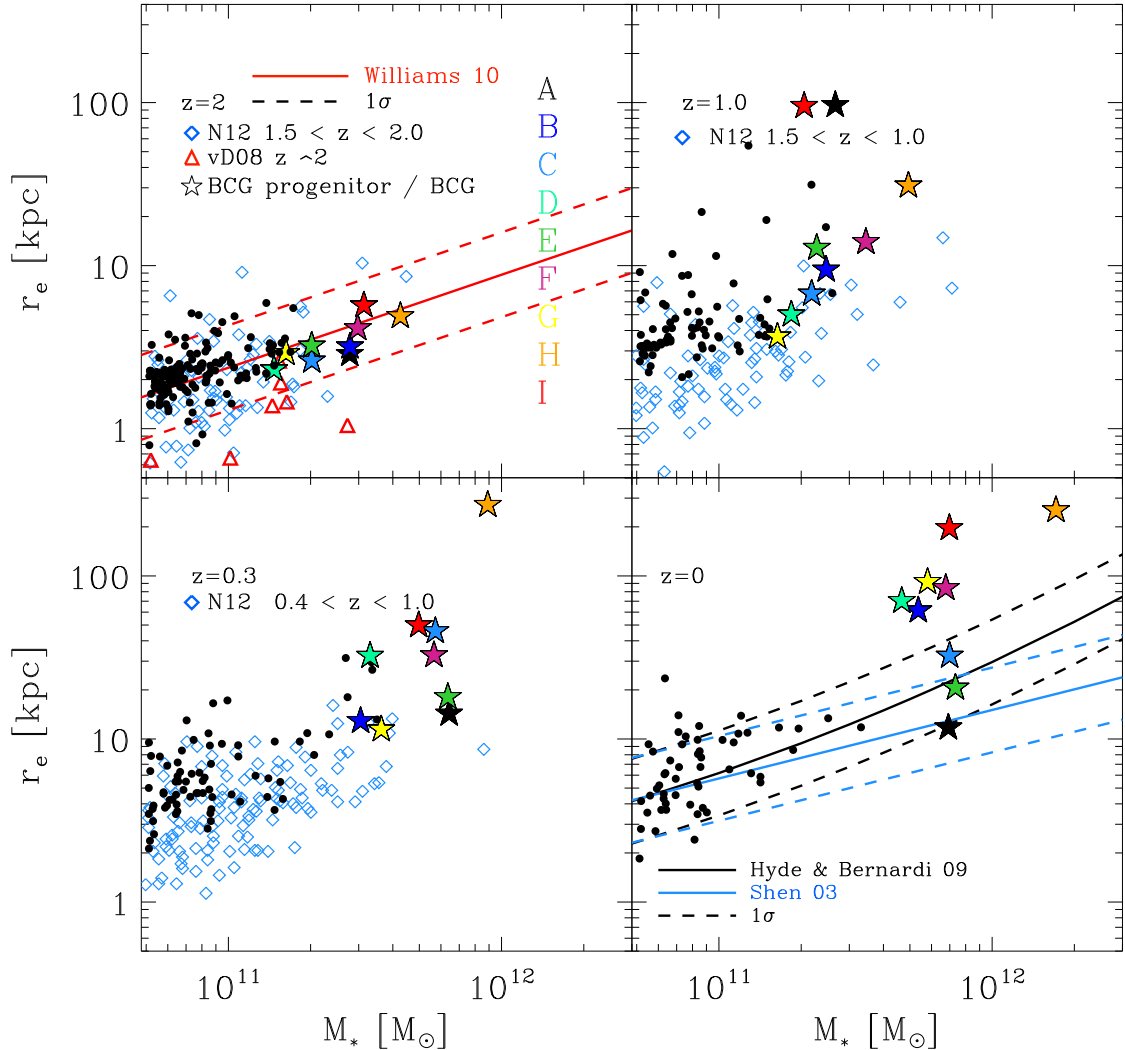}
\caption{Mass-size evolution for cluster galaxies. Note that we use a 3D definition of half-mass radius for our simulated galaxies and we compare these to de-projected half-mass radii for observations. Black dots represent simulated ellipticals, stars depict the $z=0$ BCGs and their most massive progenitors in the higher redshift panels. {\it Top left}: $z=2$ initial conditions overlaid with data from Newman et al. 2012 (N12) in blue diamonds, van Dokkum et al. (2008) in red triangles and the relation from Williams et al. 2010. {\it Top right}: $z=1.0$ mass size relation with data from N12 in the redshift range $1.5<z<1.0$. {\it Bottom left}: $z=0.3$ mass-size relation with data from N12 in the redshift range $0.4 < z <1.0$. {\it Bottom right}: $z=0$ mass-size relation overlaid with the mass-size relations of Hyde \& Bernardi (2009) and Shen et al. (2003)}
\end{figure*}

\begin{table}
 \centering
 \begin{minipage}{130mm}
  \begin{tabular}{@{}llrrrrlrlr@{}}
  \hline
Run & $m_{0}$& $m_{\mathrm{acc}}$ & $m_{\mathrm{merger}}$ & $f_{\mathrm{diffuse}}$ & $M_{\mathrm{200}}$\\

 &$ 10^{11} \rm{M_{\odot}}$& $ 10^{11} \rm{M_{\odot}}$ & $ 10^{11} \rm{M_{\odot}}$ & & $ 10^{14} \rm{M_{\odot}}$ \\
  \hline
Ph-A-2 &  4.4 & 8.6  & 7.7  &   0.07 & 8.9 \\
Ph-B-2 &  2.7 & 7.4  & 5.7  &   0.16 & 11.3 \\
Ph-C-2 &  3.2 & 7.5  & 7.0  &   0.05 & 7.5 \\
Ph-D-2 &  1.7 & 3.8  & 3.4  &   0.07 & 8.5 \\
Ph-E-2 &  2.2 & 9.0  & 7.4  &   0.14 & 8.2 \\
Ph-F-2 &  4.7 & 5.6  & 3.5  &   0.20 & 10.9 \\
Ph-G-2 &  2.5 & 5.8  & 3.1  &   0.32 & 15.8\\
Ph-H-2 &  3.6 & 3.5  & 1.4  &   0.30 & 15.6\\
Ph-I-2 &  6.6 & 16.4 & 11.7 &   0.20 & 33.0\\
\hline
\end{tabular}
\end{minipage}
\caption{Table summarising initial mass of the most massive BCG progenitor, stellar mass accreted over the last 10 Gyr, stellar contributed by mergers, the fraction of the final stellar mass contributed by the disruption of satellites (not defined as mergers). Note that we define a merger to occur when more than 50 percent of the stellar mass of a galaxy goes into the final BCG.}
\end{table}

\begin{table}
 \centering
 \begin{minipage}{130mm}
  \begin{tabular}{@{}llrrrrlrlr@{}}
  \hline
 %Run & 1:1 & 1:2 & 1:3 & 1:5 &1:10 & 1:100\\
Run & 1:1-2 & 1:3-5 & 1:10-100\\
  \hline
%Ph-A-2 & 1 & 1 & 0 & 1 & 0 & 4 \\
Ph-A-2 & 2 & 1 & 4\\
%Ph-B-2 & 0 & 1 & 4 & 2 & 2 & 2\\
Ph-B-2 & 1 & 6 & 4\\
%Ph-C-2 & 0 & 1 & 4 & 1 & 2 & 3\\
Ph-C-2 & 1 & 5 & 5\\
%Ph-D-2 & 1 & 0 & 2 & 1 & 0 & 4\\
Ph-D-2 & 1 & 3 & 5\\
%Ph-E-2 & 2 & 1 & 0 & 0 & 3 & 0\\
Ph-E-2 & 3 & 0 & 3\\
%Ph-F-2 & 0 & 1 & 0 & 0 & 2 & 0\\ 
Ph-F-2 & 1 & 0 & 2\\
%Ph-G-2 & 1 & 0 & 0 & 0 & 3 & 5\\
Ph-G-2 & 1 & 0 & 8\\
%Ph-H-2 & 0 & 0 & 0 & 1 & 1 & 1\\
Ph-H-2 & 0 & 1 & 2\\
%Ph-I-2 &  0 & 0 & 2 & 4 & 2 & 1\\
Ph-I-2 & 0 & 6 & 3\\
\hline
\end{tabular}
\end{minipage}
\caption{Table summarising the merger ratios for each BCG. Note that we define a merger when more than 50 percent of the stellar mass of a galaxy goes into the BCG.}
\end{table}

\section{Discussion}

We now turn to compare our work with previous simulation studies of BCGs. Recently, \cite{Martizzi2011} presented hydrodynamical simulations of BCGs in which the central galaxy was brought to agree with the local AMRs through the inclusion of AGN feedback. However, the surface brightness profile of the final galaxy contains a stellar core of 10 kpc. Such large stellar cores are clearly absent in real BCGs, suggesting that the AGN feedback mechanisms implemented in these simulations are too strong. While it seems that AGN feedback is the most promising way of suppressing over-cooling and star formation at the centre of clusters \citep{Croton2006,Sijacki2006,Puchwein2010}, its impact on the structure of the cluster central galaxy is probably negligible because too little gas is present in the star-dominated regions. Our work shows that BCGs can form through dissipationless mergers of galaxies between $z=2$ and $z=0$ with properties closely resembling those seen in the local Universe (surface brightness profiles, stellar-to-dark matter ratios). 

%This is re-assuring and puts important constraints for future work interpreting the result from hydrodynamical simulations. % TO BE INCLUDED OR NOT????
%We confirm the result of \cite{Ruszkowski2009} that BCGs experience many more mergers than the second brightest cluster galaxy and this explains their large sizes with a larger set of cluster simulations. However, contrary to \cite{Ruszkowski2009} not all our BCGs end up off the mass-size relation. 
%We note that these authors compared their results with the \cite{Shen2003} relation which has a shallower slope than that of \cite{Hyde2009} which we have used. Even so, when using the \cite{Shen2003} relation, we see from Figure 4 that the BCG in Ph-A is still in agreement with the normal elliptical mass-size relation. 
%It should be noted that our sample of nine clusters is rather scarce over the halo masses considered ($5\times10^{14} - 3\times10^{15}\rm{M_{\odot}}$) in order to make a firm claim on the special properties of BCGs compared to the bulk of ellipticals. To address this question would require a sample of 100 cluster simulations. 

Turning to the evolution of our by definition ``quiescent'' galaxies, we observe an evolutionary trend close to those found in observations (e.g. \cite{Newman2012}). However we caution that perhaps our experiment is not suited for such a direct comparison to observations. Indeed, the $z=2$ Universe is diverse, containing extended star-forming galaxies as well as compact galaxies, both quiescent and star forming (e.g. Figure 2 in \cite{Szomoru2012}). In this paper we have assumed that all haloes at $z=2$ contain compact quiescent galaxies. On the other hand, we can affirm that the BCGs could well be descendants of quiescent massive ellipticals at $z=2$. The best evidence for this in our experiment is that these galaxies started on the same mass-size relation as the bulk of galaxies. The high number of mergers they experience turns them into present-day merger remnants which closely match the surface brightness profiles of known BCGs.

One caveat of our experiment is that we omit the effect of the baryons on the total potential of galaxies at $z=2$. We have considered the potential of the dark-matter only simulation to represent that of $z=2$ galaxies. Including the self-gravity of the stars would in principle deepen the potential wells of the galaxies making tidal stripping of the inner-regions of haloes less efficient. However, the inclusion or omission of contraction of the dark matter haloes due to the presence of a stellar component in these simulations appears to make relatively little difference to the final results on the evolution of the BCG \citep{Ruszkowski2009, Laporte2012}.  Once more information is available on the internal structure of galaxies at $z=2$, it will be necessary to consider more complex  modelling including the stars explicitly. 
%In this sense, while our experiment differs from traditional approaches, it still improves on what has been the norm as we have to simultaneously represented a galaxy population consistent with the luminosity function in $\Lambda$CDM and the mass-size relation of quiescent galaxies at $z=2$ in our initial conditions. %This work is a first attempt towards understanding the evolution of BCGs from progenitors sharing the properties close to the observations of massive galaxies at $z=2$

\section{Conclusions}

We have studied the late formation and assembly of BCGs in the $\Lambda$CDM cosmology starting from a $z=2$ a population of ``quiescent'' galaxies resembling those observed. We did so by representing the stellar components of galaxies inside their dark matter haloes through a distribution function weighting scheme. In contrast with previous studies, our experiment {\it simultaneously} represents the luminosity function at $z=2$ as derived from abundance matching relations and the observed mass-size relation at $z=2$. Previous studies N-body studies of this have asssumed substantially too much stellar mass for a given halo mass \citep{Ruszkowski2009, Rudick2011}, thus adopting initial conditions inconsistent with modern observations and a $\Lambda$CDM cosmology.

Under the assumptions of our experiment, we predict present-day BCGs with stellar masses in good agreement with those inferred from the $z=0$ stellar-to-halo mass relation. Moreover, the surface brightness profiles of our simulated BCGs match local observations of known BCGs both in shape and in normalisation. This suggests that most BCGs have evolved passively from $z=2$ to $z=0$, forming out of pre-existing a population of compact galaxies with very little star formation after $z=2$. %We take this as compelling evidence that Brightest Cluster Galaxies in LCDM are {\it not} primordially special objects compared to the average elliptical galaxy population. The fact they show signs of peculiarity is only reminescent of the hierarchical growth of structure in which galaxy clusters form.

The large masses and sizes of cD galaxies reflect the hierarchical growth of structure in $\Lambda$CDM and their special location at the centres of galaxy clusters. Our experiment also reproduces surface brightness profiles of some of the five $z\sim 1$ BCGs in \cite{Stott2011} suggesting that these do not conflict with $\Lambda$CDM expectations. In fact, we show that our results agree well with the recent study of Lidman et al. (2012). Our simulated galaxies suggest that estimates based on fluxes within 30 kpc apertures may in some cases substantially underestimate the stellar masses of BCGs.

The Phoenix project offers an important window to study various aspects of cluster dynamics and the evolution of galaxies. Our presented scheme could be used in the future with the aim to address questions about colour and metallicity structure in BCGs using newly developed semi-analytic models (Yates et al. in prep).

%There is much work to be done to understand the evolution of the most massive ellipticals in the Universe. While surveys such as the SDSS have opened a new door to extra-galactic astronomy in this field \citep{vonderLinden2007} many authors have noticed that the quality of the photometry is not always suited to study the mass-size relation of massive ellipticals \citep{Hyde2009, Graham2008}. It is now necessary to push forward survey programs dedicated to a more careful analysis of the light profile of BCGs following the works of \cite{Graham1996, Gonzalez2005, Stott2011} in order to acquire a better picture on the evolution of BCGs with redshift. Hopefully, when the next generation telescopes will be operational such an entreprise using $10$m class telescopes will bourgeon, marking a new era for survey astronomy.

% MORE OF A DISCUSSION BUSINESS Many of the compact galaxies evolve through the accretion of smaller members and their sizes evolves predominantly through minor mergers. Recently there were some concerns with the amount by which galaxies in numerical simulation grew (e.g. Oser 2012). Our results show that even though too many stars were forming in hydrodynamical simulations, still the mechanism is viable. From our side of the experiment we are dealing with galaxies that have total potentials which are somewhat shallower than what would perhaps occur in nature (although this is a subject of its own which would owe further experimentation and understanding).

\section*{Acknowledgments}
This work was supported by the Marie Curie Initial Training Network CosmoComp (PITN-GA-2009-238356). Phoenix is a project of the Virgo Consortium. Most simulations were carried out on the Lenova Deepcomp7000 supercomputer of the super 
Computing Centre of Chinese Academy of Sciences, Beijing, China. LG 
acknowledges support from the one-hundred-talents program of the Chinese 
academy of science (CAS), the National Basic Research Program of China 
(program 973 under grant No. 2009CB24901), MPG partner Group family, 
NSFC grants (Nos. 11133003) and an STFC Advanced Fellowship. CFPL acknowledges useful discussions with Jerry Ostriker, Priya Natarajan, Laura Sales, Sadegh Khochfar and Charlie Conroy.
\bibliographystyle{mn2e}
\bibliography{master2.bib}{}
\label{lastpage}
\end{document}